\documentclass[12pt]{iopart}

\usepackage{graphicx}
\usepackage{hyperref}
\usepackage{amssymb}
\usepackage{color}
\usepackage{ulem}

\DeclareGraphicsExtensions{.eps}

\def\gz{\ifmmode{Z\hskip -4.8pt Z}
    \else{\hbox{$Z\hskip -4.8pt Z$}}\fi}

\newcommand{\be}{\begin{equation}}
\newcommand{\ee}{\end{equation}}
\newcommand{\bea}{\begin{eqnarray}}
\newcommand{\eea}{\end{eqnarray}}

\begin{document}

\title{Nonequilibrium conductance of a nanodevice for small bias voltage}
\author{A.~A.~Aligia}
\address{Centro At\'{o}mico Bariloche and Instituto Balseiro, Comisi\'{o}n Nacional
de Energ\'{\i}a At\'{o}mica, 8400 Bariloche, Argentina}
\date{\today }

\begin{abstract}
Using non-equilibrium renormalized perturbation theory, we calculate the
retarded and lesser self energies, the spectral density $\rho(\omega)$ near
the Fermi energy, and the conductance $G$ through a quantum dot as a
function of a small bias voltage $V$, in the general case of electron-hole
asymmetry and intermediate valence. The linear terms in $\omega$ and $V$ are
given exactly in terms of thermodynamic quantities. When the energy
necessary to add the first electron ($E_d$) and the second one ($E_d+U$) in
the quantum dot are not symmetrically placed around the Fermi level, $G$ has
a linear term in $V$ if in addition either the voltage drop or the coupling
to the leads is not symmetric. 
The effects of temperature are discussed. The results simplify for
a symmetric voltage drop, a situation usual in experiment.
\end{abstract}

\date{\today}
\maketitle

\bigskip

\section{Introduction}

\begin{figure}[tbp]
\includegraphics[width=8.cm]{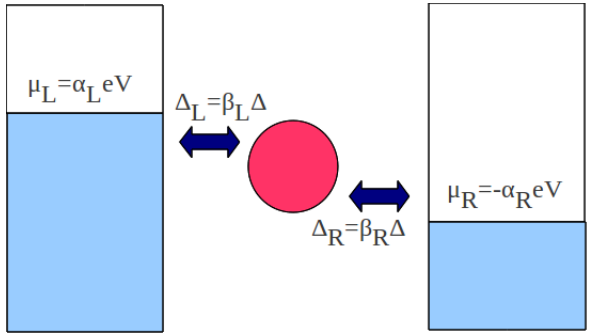}
\caption{(Color online) Scheme of the system. A QD (circle) is connected to the left ($L$) and right ($R$) 
conducting leads, with a difference $eV$ between the corresponding chemical potentials. 
Each lead hybridizes with the dot level providing a contribution $\beta_\nu \Delta$ to the resonant width 
$\Delta$.}
\label{scheme}
\end{figure}

Recently, there has been great interest in the conductance  through one
quantum dot (QD) for low applied bias voltage  and temperature, searching
for universal scaling properties \cite{grobis,scott,rinc,sela,roura,bal,kret}. In 
experiments \cite{grobis,scott,kret,gold} a semiconductor QD or a single molecule
is attached to two conducting leads (left $L$ and right $R$), a bias voltage 
$V$ is applied between the leads (see Fig.  \ref{scheme}), and the conductance $G=dI/dV$, where $I$
is the current, is measured. For $V=0$, the temperature dependence of the
conductance has been found to be very well described by the same universal
function $G(T/T_{K})$, where $T_{K}$ is the Kondo temperature, even for
systems with very different $T_{K}$ \cite{grobis,gold}. This scaling law had
been obtained from numerical renormalization group (NRG) calculations of the
impurity Anderson model \cite{chz} in the Kondo regime ($-E_{d}\gg \Delta $
and $E_{d}+U\gg \Delta $, where the Fermi energy is set as 0, $E_{d}$ is the
on-site energy, $U$ the Coulomb repulsion and $\Delta $ the resonant level
width).

In the non equilibrium situation $V\neq 0$, the problem is much tougher
theoretically. 
In this situation, exact Bethe ansatz results are available only for a simpler 
problem (the interacting resonance level model) \cite{met}  while NRG methods have serious limitations \cite{ro1}.
In some works, the spectral density $\rho_{\sigma}(\omega)$ calculated with NRG at equilibrium 
is used to calculate non-equilibrium properties, assuming that $\rho_{\sigma}(\omega)$
is not affected very much by an applied bias voltage $V$ \cite{heur,logan,serge1,cor}.
However, this approach misses the effects of broadening of $\rho_{\sigma}(\omega)$ caused
by $V$, and as a consequence, the results are quantitatively
and in some cases even qualitatively different than the correct ones \cite{st}.
     
In the Kondo limit, for either $eV\gg kT_{K}$, or $\mu
_{B}B\gg kT_{K}$ in presence of an applied magnetic field $B$, have been
determined using perturbation theory and poor man's scaling \cite{rosch}.
Using a Fermi liquid approach, based on perturbation theory in $U$ 
(PTU \cite{yam,yoshi,hor1,hor2}), and Ward identities, Oguri has determined exactly
the scaling up to second order in $kT$ and $eV$ for the symmetric Anderson
model \cite{ogu1,ogu2}. Further work considered the effect of higher order
contributions using different approximations, like PTU \cite{rinc},
non-crossing approximation \cite{roura}, or decoupling of equations of motion \cite{bal}.

The effect of asymmetric coupling to the leads $\Delta _{L}\neq \Delta _{R}$
[defined by Eq. (\ref{del})], and asymmetric drop in the bias voltage [$%
\alpha _{L}\neq \alpha _{R}$, see Eq. (\ref{mu})] has been calculated up to
second order in $T$ and $V$ using Fermi liquid approaches, for the symmetric
Anderson model ($-E_{d}=E_{d}+U$) \cite{rinc,sela}. In other works
electron-hole asymmetry was present for $V=0$ but symmetric voltage drop and
coupling to the leads was assumed \cite{roura,bal}. In all these works, the
linear terms in $T$ and $V$ are absent, and the conductance is maximum for $T=V=0$. 
In this work, we calculate exactly the linear term in $V$ in terms of
thermodynamic quantities at equilibrium.
This term is non vanishing if $-E_{d} \neq E_{d}+U
$ and in addition either $\Delta _{L}\neq\Delta _{R}$ or $\alpha _{L}\neq
\alpha _{R}$. Previous work for the infinite $U$ Anderson model, a $1/N$
expansion \cite{rati} predicts that the conductance has a linear term in $V$
of the form $2(\Delta _{R}-\Delta _{L})eV/(\Delta kT_{K})$ for $\alpha
_{L}=\alpha _{R}$.

A comparison of recent theoretical studies of the scaling properties \cite{rinc,sela,roura}
with experiment indicates that some degree of
intermediate valence (a deviation from the Kondo regime) is necessary to
explain the experimental data of Grobis {\it et al. } \cite{grobis}, in
which the effect of $V$ to decrease the conductance is about 2/3 smaller
than that expected in the Kondo regime. A good
agreement with this experiment is obtained using the Anderson model \cite{rinc,roura}, 
while the experimental results of Scott {\it et al., } \cite{scott} 
seem to be out of the reach of the model (see Section \ref{disc}).
For semiconductor QD's, it is believed that the Coulomb repulsion $U$ is not
too large in comparison with the resonant level width $\Delta =\Delta
_{R}+\Delta _{L}$ \cite{schm}. In a recent experiment in which the shot noise
is measured, these parameters were estimated as $U=0.56$ and $\Delta =0.17$ meV respectively \cite{yama}.

The above facts suggest that the expansion parameter in PTU $U/\pi \Delta$
is of order one, and then PTU might be a useful approach. In fact, PTU up to
second order in $U$ (using an interpolative scheme to recover the atomic
limit \cite{levy,kaju}) has been shown to describe well the equilibrium
conductance for $U/\pi \Delta \lesssim 2$ \cite{pro}. The results agree with
those obtained using the finite temperature density matrix renormalization
group method \cite{maru}. Interpolative PTU has also been used to calculate
persistent currents in rings with an embedded QD or side coupled to a QD \cite{pc}, 
giving results in agreement with exact calculations in small
systems \cite{pc}, and renormalization group results \cite{simon} for $U=6.25\Delta $. 
Other applications at equilibrium include impurities in
quantum corrals \cite{revi,mir}, magnetoconductance \cite{lady}, and
conductance through Hubbard chains \cite{ogu3}.

The extension of PTU up to second order to the non equilibrium (finite $V$)
case has been first considered by Hershfield {\it et al., } \cite{hersh}.
They found that for finite $V$, the current is conserved only in the
electron-hole symmetric Anderson model. Different self-consistent approaches
were proposed to overcome this shortcoming, by a suitable election of the
unperturbed Hamiltonian \cite{levy,none}. While these approaches work well in
absence of a magnetic field $B$, numerical difficulties were found for small
non-vanishing $B$ and $V$ \cite{none}. Applications of PTU up to forth order
(which seem necessary to obtain a splitting of the spectral density for
large enough $V$) \cite{fuj2,hama}, or calculations of the noise spectrum 
\cite{hama2,saka} were restricted to the symmetric model. We show in this
work that the current is conserved up to terms of order $V^{3}$ without the
need of adjusting the unperturbed Hamiltonian. We use renormalized PTU
(RPTU) to calculate the conductance for low $V$. The basic idea of RPTU as
developed by Hewson \cite{he1} is to reorganize the PTU in terms of fully
dressed quasiparticles in a Fermi liquid picture. The parameters of the
original model are renormalized and their values can be calculated exactly
from Bethe ansatz results \cite{zpb,andr,tsve,bet,pedro}, or accurately 
using NRG \cite{he2,kris1,kris2,hbo,bulla}. 
One of the main advantages is that the renormalized
expansion parameter $\widetilde{U}/(\pi \widetilde{\Delta })\eqslantless 1$,
being 1 in the extreme Kondo regime ($U\rightarrow \infty $). Here we assume 
$B=0$. Exact results for small magnetic field were obtained using RPTU in
the symmetric case \cite{hbo}, while for large $B$, calculations with
interpolative PTU were presented \cite{none}.

We calculate the self energies and the spectral density near the Fermi
energy for small values of the frequency $\omega $ and bias voltage $V$, and
the conductance for small $V$. The linear terms in $\omega $ and $V$ are
given exactly in terms of the occupation at the dot, magnetic susceptibility and 
specific heat at equilibrium. The lesser self energy
is given exactly up to quadratic terms. We also show that the current is
conserved up to terms of order $V^{3}$ in PTU. 

The paper is organized as
follows. In Section II we describe the system and the impurity Anderson
model used to represent it. We also review briefly the formalism of the
perturbation theory and the idea of RPTU. Section III contains the results
of the calculations, and several limiting cases of interest in which the
general expressions become simpler. Section \ref{disc} contains a summary
and a discussion.

\section{Model and formalism}

\subsection{Model}
A scheme of the model is displayed in Fig. \ref{scheme}.
The QD interacting with two conducting leads is described by the spin 1/2
Anderson model. In general, to use PTU, it is convenient to split the
Hamiltonian into a noninteracting part $H_{0}$ and a perturbation $H^{\prime
}$ as%
\begin{eqnarray}
H &=&H_{0}+H^{\prime },  \nonumber \\
H_{0} &=&\sum_{k\nu \sigma }\varepsilon _{k\nu }\,c_{k\nu \sigma }^{\dagger
}c_{k\nu \sigma }+\sum_{\sigma }\varepsilon _{eff}^{\sigma }\,n_{d\sigma } 
\nonumber \\
&&+\sum_{k\nu \sigma }\left( V_{k\nu }\,c_{k\nu \sigma }^{\dagger }d_{\sigma
}+{\rm H.c.}\right) ,  \nonumber \\
H^{\prime } &=&\sum_{\sigma }\left( E_{d}-\varepsilon _{eff}^{\sigma
}\right) \,n_{d\sigma }+U\,n_{d\uparrow }n_{d\downarrow },  \label{h}
\end{eqnarray}%
where $n_{d\sigma }=d_{\sigma }^{\dagger }d_{\sigma }$, and $\nu =L,R$
refers to the left and right leads, with chemical potentials

\begin{equation}
\mu _{L}=\alpha _{L}eV,~~\mu _{R}=-\alpha _{R}eV.  \label{mu}
\end{equation}%
respectively, with $\alpha _{L}+\alpha _{R}=1$. Similarly, the couplings to
the leads assumed independent of frequency are expressed in terms of the
total resonant level width $\Delta =\Delta _{L}+\Delta _{R}$ as

\begin{equation}
\Delta _{\nu }=\pi \sum_{k}|V_{k\nu }|^{2}\delta (\omega -\varepsilon _{k\nu
})=\beta _{\nu }\Delta .  \label{del}
\end{equation}

In general $\varepsilon _{eff}^{\sigma }$ is determined selfconsistently,
except for the electron-hole symmetric case with $B=0$, for which $\varepsilon _{eff}^{\sigma }=0$ \cite{levy,none}.

\subsection{Green's functions and self energies}

The one-particle properties of the system, including the current, are
determined by three types of independent one-particle Green's functions, the retarded 
$G^{r}(\omega )$, the advanced $G^{a}(\omega )$ which are the complex
conjugate of $G^{r}(\omega )$, and the lesser ones $G^{<}(\omega )$. The
retarded Green's function of the electrons at the dot for spin $\sigma $,
can be written as 

\begin{equation}
G_{d\sigma }^{r}(\omega )=\frac{1}{\omega -\varepsilon _{eff}^{\sigma
}+i\Delta -\Sigma _{\sigma }^{r}(\omega )}.  \label{gr}
\end{equation}%
In PTU up to second order in $U$, the retarded self energy is approximated as \cite{none}

\begin{equation}
\Sigma _{\sigma }^{r}(\omega )=E_{d}^{\sigma }-\varepsilon _{eff}^{\sigma
}+U\langle n_{d\overline{\sigma }}\rangle +\Sigma _{\sigma }^{r2}(\omega ),
\label{sr}
\end{equation}%
where $\Sigma _{\sigma }^{r2}$ contains the contribution of order $U^{2}$.

The lesser Green's function can be written in the form \cite{none}

\begin{equation}
G_{d\sigma }^{<}(\omega )=|G_{d\sigma }^{r}(\omega )|^{2}\left( \frac{%
g_{d\sigma }^{<}(\omega )}{|g_{d\sigma }^{r}(\omega )|^{2}}-\Sigma _{\sigma
}^{<}(\omega )\right) ,  \label{gl}
\end{equation}%
where $g_{d\sigma }^{r}(\omega )$, $g_{d\sigma }^{<}(\omega )$ are the
retarded and lesser noninteracting Green's functions, and $\Sigma _{\sigma
}^{<}(\omega )$ is the lesser self energy.

For large values of $U/\Delta $, ordinary PTU in $U$ is not reliable and
Hewson proposed to reorganize the perturbation series in terms of
renormalized parameters (which we denote with a tilde) such that 
$\widetilde{U}/(\pi \widetilde{\Delta })\eqslantless 1$. Within RPTU, the low frequency
part of $G_{d\sigma }^{r}(\omega )$ is approximated as \cite{he1}

\begin{equation}
G_{d\sigma }^{r}(\omega )\simeq \frac{z}{\omega -\widetilde{\varepsilon }_{eff}^{\sigma }+i\widetilde{\Delta }
-\widetilde{\Sigma }_{\sigma
}^{r}(\omega )},  \label{gra}
\end{equation}
where

\begin{eqnarray}
z &=&[1-\partial \Sigma _{\sigma }^{r}/\partial \omega ]^{-1},~~
\widetilde{\varepsilon }_{eff}^{\sigma }=z[\varepsilon _{eff}^{\sigma
}+\Sigma _{\sigma }^{r}(0)],  \nonumber \\
\widetilde{\Delta } &=&z\Delta , ~~ \widetilde{\Sigma }_{\sigma
}^{r}(\omega )=z\Sigma _{\sigma }^{{\rm rem}}(\omega ),  \label{z}
\end{eqnarray}%
and the remainder retarded self-energy is defined as

\begin{equation}
\Sigma _{\sigma }^{{\rm rem}}(\omega )=\Sigma _{\sigma }^{r}(\omega
)-\Sigma _{\sigma }^{r}(0)-\omega \partial \Sigma _{\sigma }^{r}/\partial
\omega .  \label{rem}
\end{equation}%
In Eqs. (\ref{gra}) and (\ref{rem}), $\Sigma _{\sigma }^{r}(0)$ and $%
\partial \Sigma _{\sigma }^{r}/\partial \omega $ are evaluated at $\omega
=T=V=0$.

\begin{figure}[tbp]
\includegraphics[width=8.cm]{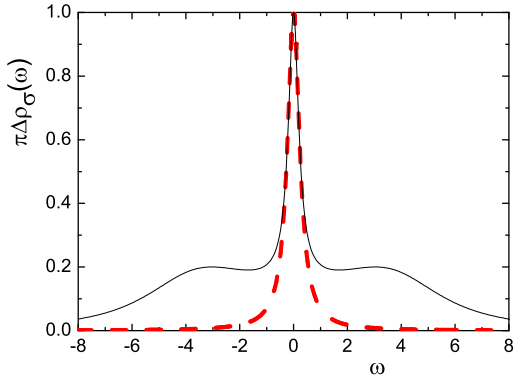}
\caption{(Color online) Full line: spectral density of the Anderson model
for $U=7 \Delta$ and $E_d=-U/2$. Dashed line: corresponding result for
non-interacting quasiparticles (see text).}
\label{f1}
\end{figure}

In Fig. \ref{f1} we compare the spectral density 
$\rho _{\sigma }(\omega )=-$Im$G_{d\sigma }^{r}(\omega )/\pi $ evaluated within PTU up to second order with
the corresponding noninteracting renormalized result 
($z\widetilde{\rho }_{0}(\omega )=-$Im$\widetilde{G}_{d\sigma }^{r}(\omega )/\pi $ with 
$\widetilde{\Sigma }_{\sigma }(\omega )=0$) obtained from Eqs. (\ref{gra}) and (\ref{z}) for
the symmetric Anderson model ($\widetilde{\varepsilon }_{eff}^{\sigma }=0$).
We obtain $z=0.2744$ for the parameters of the figure.
This comparison has been made by Rejec and Ram\v{s}ak for other parameters \cite{rej}.
One can see that already non-interacting quasiparticles reproduce rather
well the low-energy behavior. A similar comparison, for a case with
non-trivial frequency dependent $\Delta (\omega )$ has also been 
made \cite{vaug}. showing also a very good agreement for low $|\omega |$. An accurate
calculation of $\widetilde{\Sigma }_{\sigma }(\omega )$ including high orders
in the perturbation expansion, should lead in turn to a precise description of the spectral 
density at high energies, since nothing is lost in the reorganization of the
perturbative expansion. However, the calculation of higher order terms
becomes involved due to the need of considering counterterms and corrections
to the parameters \cite{he1}.

The spectral density evaluated at the Fermi energy is related with the
occupancy by the Friedel sum rule \cite{lan}

\begin{equation}
\rho _{\sigma }(0)=\frac{\sin ^{2}(\pi \langle n_{d\sigma }\rangle )}{\pi
\Delta },  \label{rho0}
\end{equation}%
which allows one to relate the effective dot level with its occupancy, using Eq.
(\ref{gra})

\begin{equation}
\widetilde{\varepsilon }_{eff}^{\sigma }=\widetilde{\Delta }\cot (\pi
\langle n_{d\sigma }\rangle ).  \label{eff}
\end{equation}

$\widetilde{\Sigma }_{\sigma }^{{\rm rem}}(\omega )$ can be calculated
perturbatively using a renormalized interaction $\widetilde{U}$ which is
given by the vertex function $\Gamma $ at the Fermi energy \cite{he1,he2}. In
absence of magnetic field, the subscript $\sigma $ can be dropped. The
linear term in the specific heat and the impurity contribution to the
magnetic susceptibility (determined using Ward identities) are given by \cite{he1}

\begin{eqnarray}
\gamma_{C} &=& 2\pi ^{2}k\widetilde{\rho }_{0}(0)/3,  \label{gam} \\
\chi &=& (g\mu _{B})^{2}(1+\widetilde{U}\widetilde{\rho }_{0}(0))/2,
\label{xi}
\end{eqnarray}%
where

\begin{equation}
\widetilde{\rho }_{0}(\omega )=\frac{\widetilde{\Delta }/\pi }{(\omega -%
\widetilde{\varepsilon }_{eff})^{2}+\widetilde{\Delta }^{2}}  \label{rhor}
\end{equation}%
is the density of free renormalized quasiparticles. Eqs. (\ref{rho0}), (\ref%
{gam}), (\ref{xi}) and an accurate knowledge of thermodynamic quantities
from Bethe ansatz or NRG, permits a precise determination of $z=\rho (0)/%
\widetilde{\rho }_{0}(0)$, and the renormalized interaction through the
Wilson ratio $R=1+\widetilde{U}\widetilde{\rho }_{0}(0)$.

To order $\widetilde{U}^{2}$, the renormalized self energies can be written
as \cite{none}

\begin{eqnarray}
\widetilde{\Sigma }^{r}(\omega ) &=&\widetilde{U}^{2}\int d\epsilon
_{1}d\epsilon _{2}d\epsilon _{3}\frac{\widetilde{\rho }_{0}(\epsilon _{1})%
\widetilde{\rho }_{0}(\epsilon _{2})\widetilde{\rho }_{0}(\epsilon _{3})}{%
\omega +\epsilon _{3}-\epsilon _{1}-\epsilon _{2}+i\eta }  \nonumber \\
&&\times \lbrack (1-\tilde{f}(\epsilon _{1}))(1-\tilde{f}(\epsilon _{2}))%
\tilde{f}(\epsilon _{3})  \nonumber \\
&&+\tilde{f}(\epsilon _{1})\tilde{f}(\epsilon _{2})(1-\tilde{f}(\epsilon
_{3}))],  \label{srro}
\end{eqnarray}

\begin{eqnarray}
\tilde{\Sigma}^{<}(\omega ) &=&z\Sigma ^{<}(\omega )  \nonumber \\
&=&-2i\pi \widetilde{U}^{2}\int d\epsilon _{1}d\epsilon _{2}\widetilde{\rho }%
_{0}(\epsilon _{1})\widetilde{\rho }_{0}(\epsilon _{2})\widetilde{\rho }%
_{0}(\epsilon _{1}+\epsilon _{2}-\omega )  \nonumber \\
&&\times \tilde{f}(\epsilon _{1})\tilde{f}(\epsilon _{2})(1-\tilde{f}%
(\epsilon _{1}+\epsilon _{2}-\omega )),  \label{slro}
\end{eqnarray}%
where $\eta $ is a positive infinitesimal and $\tilde{f}(\omega )=\sum_{\nu
}\beta _{\nu }(\omega )f(\omega -\mu _{\nu })$ is an average of the Fermi
functions at the two leads $\left[ f(\omega )=1/(e^{\omega /kT}+1)\right] $.

\subsection{The current}

Using the Keldysh formalism \cite{past,meir}, it has been shown that the
current flowing between the left lead and the dot is

\begin{equation}
I_{L}=\frac{4ie\Delta _{L}}{h}\int d\omega \left[ 2if(\omega -\mu _{L}) 
{\rm Im} G_{d}^{r}(\omega )+G_{d}^{<}(\omega )\right].  \label{il}
\end{equation}%
Similarly, the current with spin $\sigma $ flowing between the dot and the
right lead is

\begin{equation}
I_{R}=-\frac{4ie\Delta _{R}}{h}\int d\omega \left[ 2if(\omega -\mu _{R})%
{\rm Im} G_{d}^{r}(\omega )+G_{d}^{<}(\omega )\right].  \label{ir}
\end{equation}
Of course, since the current is conserved one should have $I_{L}=I_{R}=I$.
Then, from a weighted average of both expressions, $G_{d}^{<}$ can be
eliminated, giving \cite{meir}

\begin{equation}
I=\frac{8\pi \beta _{L}\beta _{R}e\Delta }{h}\int d\omega \rho (\omega
)[f(\omega -\mu_L)-f(\omega -\mu_R)].  \label{i}
\end{equation}

\section{Self energies, spectral density and conductance for small bias
voltage}

\subsection{Self energies}
\label{self}

It is easy to see that to evaluate the differential conductance $G=dI/dV$ at
zero temperature to order $V^{n}$, it is enough to calculate the spectral
density and self energies up to total order $n$ in $\omega $ and $V$ 
(all terms of the form $\omega^l V^{m-l}$, with $0\leq m \leq l$, $l \leq n$),    since the
effective interval of integration in Eq. (\ref{i}) is $eV$, because the
Fermi functions $f(\omega )$ become step functions $\theta (-\omega )$ at $%
T=0$. For the same reason, the intervals of the two integrations in Eq. (\ref{slro}) 
are of order $eV$, and therefore one obtains $\tilde{\Sigma}%
^{<}(\omega )$ up to terms of second order, taking $\widetilde{\rho }%
_{0}(\epsilon )$ at the Fermi energy $\epsilon =0$. Performing the
integration we find

\begin{eqnarray} 
\tilde{\Sigma}^{<}(\omega ) &=&-i\pi \lbrack \widetilde{\rho }_{0}(0)]^{3}%
\widetilde{U}^{2}  \nonumber \\
&&\times \{\theta (\alpha _{L}eV-\omega )\beta _{L}(\beta _{L}^{2}+2\beta
_{R}^{2})(\alpha _{L}eV-\omega )^{2}  \nonumber \\
&&+\theta (-\alpha _{R}eV-\omega )\beta _{R}(2\beta _{L}^{2}+\beta
_{R}^{2})(\alpha _{R}eV+\omega )^{2}  \nonumber \\
&&+\theta \lbrack (1+\alpha _{L})eV-\omega ]\beta _{L}^{2}\beta
_{R}[(1+\alpha _{L})eV-\omega ]^{2}  \nonumber \\
&&+\theta \lbrack -(1+\alpha _{R})eV-\omega ]\beta _{L}^{2}\beta
_{R}[(1+\alpha _{R})eV+\omega ]^{2}.  \nonumber \\
\label{sl}
\end{eqnarray}

Using ${\rm Im}(\omega +i\eta )^{-1}=-i\pi \delta (\omega )$ and Eq. (\ref%
{srro}) it is easy to realize that the above arguments also work for ${\rm Im}\widetilde{\Sigma }(\omega )$ 
and performing the two remaining
integrations [after the use of the delta function $\delta (\omega +\epsilon
_{3}-\epsilon _{1}-\epsilon _{2})$] one obtains up to terms of second order
in $\omega $ and $V$

\begin{eqnarray}
{\rm Im}\widetilde{\Sigma }^{r}(\omega ) &=&-\frac{\pi }{2}[\widetilde{\rho 
}_{0}(0)]^{3}\widetilde{U}^{2}[\omega ^{2}-2\gamma \omega eV+\delta (eV)], 
\nonumber \\
\gamma  &=&\alpha _{L}\beta _{L}-\alpha _{R}\beta _{R},  \nonumber \\
\delta  &=&\gamma ^{2}+3\beta _{L}\beta _{R},  \label{sret}
\end{eqnarray}%
as previously found by Oguri \cite{ogu2}.

Note that higher order terms in the perturbation series in $\widetilde{U}$
would lead to terms of higher order in $V$. Therefore, the above results are
exact. They can be used to test approximations. For example, from Eqs. (\ref{sl}) and 
(\ref{sret}), one realizes that the approximation for the lesser
self energy $\tilde{\Sigma}^{<}(\omega )\simeq 2i\tilde{f}(\omega ){\rm Im}\widetilde{\Sigma }^{r}(\omega )$ 
used in decoupling approximations \cite{note3,bal,ng}, 
is only valid for $V=0$ (or $U=0$). 

From Eqs. (\ref{gl}), (\ref{gra}), (\ref{z}), (\ref{il}) and (\ref{ir}) one
obtains

\begin{equation}
I_{L}-I_{R}=-\frac{4e}{zh}\int d\omega |G_{d}^{r}(\omega )|^{2}[2\tilde{f}%
(\omega ){\rm Im}\widetilde{\Sigma }^{r}(\omega )+i\tilde{\Sigma}%
^{<}(\omega )].  \label{dif}
\end{equation}%
The integral introduces another factor $eV$. Thus, in order to obtain $%
I_{L}-I_{R}$ to order $V^{3}$, one can replace $|G_{d}^{r}(\omega )|^{2}$ by
its value at $\omega =0$ and move it outside the integral. Using Eqs. (\ref%
{sl}) and (\ref{sret}) the integral turns out to vanish. Then $%
I_{L}-I_{R}=O(V^{4})$ and the current is conserved up to order $V^{3}$.

In contrast to $\tilde{\Sigma}^{<}(\omega )$ and ${\rm Im}\widetilde{\Sigma 
}^{r}(\omega )$, an accurate calculation of the real part of the retarded
self energy is more difficult. As it is apparent from Eq. (\ref{srro}) or
alternative expressions \cite{hersh,none}, the calculation of ${\rm Re}\widetilde{\Sigma }^{r}(\omega )$ 
involves an integration at high energies,
for which the approximation given by noninteracting quasiparticles [Eq. (\ref%
{gra}) neglecting $\widetilde{\Sigma }^{r}(\omega )$] is not good enough
(see Fig. \ref{f1}). Fortunately $\widetilde{\Sigma }^{r}(0)=0$ and $\partial 
\widetilde{\Sigma }^{r}(\omega )/\partial \omega=0$ by construction [see Eqs. (\ref{z}) and (\ref{rem})], 
while $\partial \widetilde{\Sigma }^{r}/\partial eV$ can be calculated from Ward
identities: using the results of Oguri \cite{ogu2}, one sees that $\partial 
\widetilde{\Sigma }^{r}/\partial eV=\gamma \partial \widetilde{\Sigma }%
^{r}/\partial \mu $, where $\mu $ is a shift of both chemical potentials. In
addition, using Eqs. (3.15) to (3.18) of Ref. \cite{yoshi}, 
and $\partial\widetilde{\Sigma }^{r}/\partial \omega =0$, one obtains $\partial \widetilde{\Sigma }%
^{r}/\partial \mu =\widetilde{\rho }_{0}(0)\widetilde{U}$. Combining both
results one has

\begin{equation}
\partial \widetilde{\Sigma }^{r}/\partial eV=\gamma x.  \label{dsmu}
\end{equation}%
where we call $x=\widetilde{\rho }_{0}(0)\widetilde{U}=R-1$, where $R$ is
the Wilson ratio mentioned in the previous section. Note that for $\langle
n_{d\sigma }\rangle =1/2$ and $U\rightarrow +\infty $, corresponding to the
Kondo limit, $x=1$ \cite{he1}. This means that in this limit, the position of the Kondo
resonance at equilibrium remains at the Fermi energy [Eqs. (\ref{gra}), (\ref{eff}), 
$\widetilde{\Sigma }_{\sigma }^{r}(0)=0$ and $\partial \widetilde{\Sigma }^{r}/\partial \mu =x$] 
if the chemical potential is shifted, as expected from the common wisdom on the Kondo resonance.

\subsection{Spectral density and conductance }

Using Eqs. (\ref{gra}), (\ref{rho0}), (\ref{eff}), (\ref{sret}) and (\ref%
{dsmu}) one obtains up to quadratic terms in $\omega $ and $eV$, calling for
brevity $c=\cos (\pi \langle n_{d\sigma }\rangle )$ and $s=\sin (\pi \langle
n_{d\sigma }\rangle )$

\begin{eqnarray}
\frac{\rho (\omega )}{\rho (0)} &=&1+A\frac{\omega }{\widetilde{\Delta }}+B%
\frac{eV}{\widetilde{\Delta }}+C\left( \frac{\omega }{\widetilde{\Delta }}%
\right) ^{2}  \nonumber \\
&&+D\frac{\omega eV}{\widetilde{\Delta }^{2}}+E\left( \frac{eV}{\widetilde{%
\Delta }}\right) ^{2},  \nonumber \\
A &=&2sc, ~~ B=-2\gamma xsc,  \nonumber \\
C &=&s^{2}[4c^{2}-1+x^{2}(1-2s^{2})/2]  \nonumber \\
&&-sc\widetilde{\Delta }\partial ^{2}{\rm Re}\widetilde{\Sigma }%
^{r}/\partial \omega ^{2},  \nonumber \\
D &=&\gamma xs^{2}[2-8xc^{2}-x(1-2s^{2})]  \nonumber \\
&&-2sc\widetilde{\Delta }\partial ^{2}{\rm Re}\widetilde{\Sigma }%
^{r}/\partial \omega \partial eV,  \nonumber \\
E &=&x^{2}s^{2}[\gamma ^{2}(4c^{2}-1)+\delta (1-2s^{2})/2]  \nonumber \\
&&-sc\widetilde{\Delta }\partial ^{2}{\rm Re}
\widetilde{\Sigma }^{r}/\partial (eV)^{2}.  \label{ron}
\end{eqnarray}
This equation might be regarded as extension of the Friedel sum rule Eq. 
(\ref{rho0}) to finite small frequency and bias voltage. The annoying terms
in ${\rm Re}\widetilde{\Sigma}^r$ enter only the second order terms and
vanish for the symmetric Anderson model. Inserting the above result in Eq. 
(\ref{i}) and deriving with respect to $V$ the conductance up to terms of
order $V^{2}$ is obtained

\begin{eqnarray}
G(V) &=&G(0)\left[ 1+L\frac{eV}{\widetilde{\Delta }}`+M\left( \frac{eV}{%
\widetilde{\Delta }}\right) ^{2}\right] ,  \label{g} \\
G(0) &=&4\beta _{L}\beta _{R}s^{2}\frac{2e^{2}}{h},  \label{g0} \\
L &=&2sc(\alpha _{L}-\alpha _{R}-2\gamma x),  \label{l1} \\
M &=&C(\alpha _{L}^{3}+\alpha _{R}^{3})+3D(\alpha _{L}-\alpha _{R})/2+3E.
\label{m1}
\end{eqnarray}

This is the main result of this work. Except for the second derivatives of $%
{\rm Re}\widetilde{\Sigma }^{r}$, the remaining quantities are given
exactly in terms of the characteristic energy scale $\widetilde{\Delta }$,
the total occupation $n=2\langle n_{d\sigma }\rangle $ at the dot, the
Wilson ratio $R=x+1$, the ratio of the couplings to the leads $\beta _{\nu }$
and the distribution of the potential decays $\alpha _{\nu }$. The energy $%
\widetilde{\Delta }$ can be obtained from thermodynamic quantities. For
example from Eqs. (\ref{eff}), (\ref{gam}) and (\ref{rhor}):

\begin{equation}
\widetilde{\Delta }=\frac{2\pi k}{3\gamma _{C}}\sin ^{2}(\pi n/2).
\label{delt}
\end{equation}

Up to second order in the interaction, $\partial ^{2}{\rm Re}\widetilde{%
\Sigma }^{r}/\partial \omega ^{2}$ has been calculated by Horvati\'{c} and
Zlati\'{c} \cite{hor1,hor2}. However, for the reasons discussed above, we do
not expect this (rather complicated) result to be accurate enough for strong
interaction.

\subsection{Particular cases}

The total occupation of the dot $n$ can be controlled by the gate voltage $%
V_{g}$. Assuming that the coupling to leads is not affected by $V_{g}$, the
occupation $n=1$ maximizes $G(0)$. However, experimentally, the scaling
properties of $G(V,T)$ have been studied not only for $V_{g}$ that maximizes 
$G(0)$, but for other values as well \cite{grobis}. If one starts from the
symmetric Anderson model, then by symmetry $n=1$ for $V=0$, $c=\cos (\pi
n/2) $ vanishes and with it all terms involving second derivatives of ${\rm Re}\widetilde{\Sigma }^{r}$. 
In this case, Eq. (\ref{g}) simplifies to

\begin{eqnarray}
G(V)/G(0) &=&1-\frac{(eV)^{2}}{2\widetilde{\Delta }^{2}}[(2+x^{2})(\alpha
_{L}^{3}+\alpha _{R}^{3})  \nonumber \\
&&-3\gamma x(2+x)(\alpha _{L}-\alpha _{R})  \nonumber \\
&&+9x^{2}(\gamma ^{2}+\beta _{L}\beta _{R})],  \label{sela}
\end{eqnarray}%
which is equivalent to the result found previously by Sela and Malecki using
a different approach \cite{sela}.

To model the voltage drops, in some works \cite{rinc,schm} it has been
assumed that the shift in the average chemical potential with applied
voltage $\Delta \mu =\gamma eV=0$ [see Eq. (\ref{sret})]. However, even in
cases with very asymmetric couplings, the observed structure of the diamonds
in $G(V)$ indicates that $\alpha _{L}\simeq \alpha _{R}$ \cite{serge,note1}. Taking 
$\alpha _{L}=\alpha _{R}$, then $\gamma =(\beta _{L}-\beta _{R})/2$, and the
coefficients of Eq. (\ref{g}) simplify to

\begin{eqnarray}
L &=&-2sc(\beta _{L}-\beta _{R})x,  \nonumber \\
M &=&s^{2}[c^{2}-\frac{1}{4}+x^{2}(\frac{11}{4}-4s^{2}-6\beta _{L}\beta
_{R}c^{2})]  \nonumber \\
&&-sc\widetilde{\Delta }[\partial ^{2}{\rm Re}\widetilde{\Sigma }%
^{r}/\partial \omega ^{2}+3\partial ^{2}{\rm Re}\widetilde{\Sigma }%
^{r}/\partial (eV)^{2}].  \label{gaeq}
\end{eqnarray}%
Note that in a $1/N$ expansion of the infinite $U$ Anderson model, the value 
$L=-2(\beta _{L}-\beta _{R})$ has been found \cite{rati}.

\subsection{The maximum of $G(V)$}

In the general case for which the linear term in $V$, $M\neq 0$, the maximum
of $G(V)$ does not lie at $V=0$, but at $V_{\max }=-L/(2M)$, with $G_{\max
}/G(0)=1-L^{2}/(4M)$. We discuss the case $\alpha _{L}=\alpha _{R}$ and $%
2/3<n<4/3$ for which the curvature of the unperturbed quasiparticle density
of states is positive [see Eqs. (\ref{eff}) and (\ref{rhor})]. In this case,
because $L\sim x$, and $M<0$, the largest possible $G_{\max }$ is obtained
for high interaction $x=\widetilde{\rho }_{0}(0)\widetilde{U}$. To estimate
an upper bound for $G_{\max }$, we neglect the second derivatives 
of ${\rm Re}\widetilde{\Sigma }^{r}$ in Eqs. (\ref{gaeq}) and take the maximum
possible value of $\widetilde{U}/\pi \widetilde{\Delta }$, namely one, which
implies $x=s^{2}$ [see Eqs. (\ref{eff}) and (\ref{rhor})]. This gives

\begin{equation}
G_{\max }\leq 4Ps^{2}\frac{2e^{2}}{h}[1+\frac{4c^{2}(1-4P)}{%
16s^{4}-7s^{2}-3+24Ps^{2}c^{2}}].  \label{gmax}
\end{equation}%
where the product $P=\beta _{L}\beta _{R}$. This function always increases
with increasing $P$. Since the maximum value of $P$ is 1/4, one has more
simply

\begin{equation}
G_{\max }\leq \frac{2e^{2}}{h}\sin ^{2}(\pi n/2),  \label{gmax2}
\end{equation}%
for $\alpha _{L}=\alpha _{R}$ and $|n-1|<1/3$.

\subsection{Effect of temperature}

For $n=1$, the effects of temperature $T$ and voltage $V$ on the spectral
density and the conductance are additive up to total second order in $\omega $, $T$%
, and $V$. The corrections to the self energy and conductance for $V=0$ in
this case, were discussed previously \cite{rinc,sela,ogu1,ogu2}. In particular

\begin{eqnarray}
\frac{\rho(\omega,T)}{\rho(0,0)}&=&1-\frac{(2+x^2) \omega^2 + (\pi x kT)^{2}}
{2 \widetilde{\Delta }^{2}},
\label{rhot} \\
\frac{G(T)}{G(0)}&=&1-\frac{(\pi kT)^{2}}{3\widetilde{\Delta }^{2}}(1+2x^{2}).
\label{gt}
\end{eqnarray}
For $n \neq 1$, the corrections to the self energy in second order in $\widetilde{U}$ and 
up to second order in $\omega $ and $T$ were calculated
by Horvati\'{c} and Zlati\'{c} \cite{hor1,hor2}. There is however in general (for $V \neq 0$) a term linear in 
$T$ which comes from the term of order $\widetilde{U}$ in the self energy
[see Eq. (\ref{sr})], or $\partial {\rm Re}\widetilde{\Sigma }^{r}/\partial
T=\widetilde{U}\partial \langle n_{d\overline{\sigma }}\rangle /\partial T$.
An accurate determination of the last derivative is not possible with the
knowledge of the Green functions only for frequencies near the Fermi energy.
For $V=0$, $\partial\langle n_{d\sigma }\rangle /\partial T=0$ because the system is 
a Fermi liquid and the lowest order correction to the occupancy goes as $T^2$ \cite{zpb}.

Including only terms linear in $\omega$, $V$ and $T$, and using the results 
of the previous section one obtains 
\begin{eqnarray}
\frac{\rho(\omega)}{\rho(0)}&=&1+\frac{\sin(\pi n)}{\widetilde{\Delta }}[\omega -\gamma x eV 
-\frac{zU}{2 } \frac{\partial n^2}{\partial V \partial T}VT],
\label{rhot2} \\
\frac{G(T)}{G(0)}&=&1+\frac{\sin(\pi n)}{\widetilde{\Delta }}[eV(\alpha_L-\alpha_R -2 \gamma x) \nonumber \\
&&-e z U \frac{\partial n^2}{\partial V \partial T}VT ],
\label{gt2}
\end{eqnarray}
where the derivative is evaluated at $V=T=0$.

\section{Summary and discussion}

\label{disc}

Using Fermi liquid properties and non equilibrium renormalized perturbation
theory, we have calculated the self energies and spectral density near the Fermi energy, and the conductance
for small bias voltage $V$ compared to the characteristic energy scale $\widetilde{\Delta }$. 
We have determined exactly the linear term in $V$
in the conductance $G$, in terms of the $\widetilde{\Delta }$, 
the occupation of the dot $n$ and the Wilson ratio $R=x+1$. 
This term is different from zero if $n\neq 1$ and if in addition
either the voltage drop is asymmetric ($\alpha _{L}\neq \alpha _{R}$) or the
coupling to the leads is asymmetric ($\beta _{L}\neq \beta _{R}$). 
It is important to interpret experiments, because due to the measurement method, 
there is a small voltage offset \cite{serge}. 
For $\alpha _{L}=\alpha _{R}$, the sign of this term and its dependence on $\beta
_{L}-\beta _{R}$ agrees with previous results based on a $1/N$ expansion of
the infinite $U$ Anderson model \cite{rati}. 
In addition, the observed 
$G(V)$ is asymmetric with a $V^3$ term \cite{serge}, which is unfortunately beyond the validity 
of our approach, because the current is conserved only to order $V^3$. 

For an accurate calculation of the terms quadratic in $V$ when $n\neq 1$, it
is necessary to know second derivatives of the real part of the self energy
with respect to $V$ and frequency $\omega $. These might be calculated
combining numerical-renormalization-group (NRG) calculations \cite{bulla} with
renormalized perturbation theory \cite{ogu2,he2}. For $n \neq 1$ and finite bias voltage $V$, there is
a correction of the real part of the self energy with temperature $T$,
proportional to $\partial n / \partial T$, which is also beyond the reach of
low-energy expansions. For $V=0$, $\partial n / \partial T=0$ \cite{zpb}, but this is not 
necessarily the case out of equilibrium.

Values of the Wilson ratio $R$ for the asymmetric Anderson model can be
found for example in Table VIII of Ref. \cite{kris2}. In the extreme Kondo limit, 
$R \rightarrow 2$ ($x \rightarrow 1$) and $n \rightarrow 1$ simultaneously. 
Then, the effects of asymmetry and the coefficients of the second derivatives 
of the real part of the self energy tend to vanish.

The shape of the diamonds in experiments suggest a symmetric voltage drop 
($\alpha _{L}=\alpha _{R}$) \cite{note1}, except in arrangements like that of a scanning
tunneling microscope, for which there is a very asymmetric coupling of the
nanoscopic systems to the conductors ($\beta _{L}/\beta _{R}\gg 1$ or $\beta
_{L}/\beta _{R}\ll 1$) . In the latter case, the conductance is very small,
due to the factor $\beta _{L}\beta _{R}$ in the expression for the
conductance [see Eq. (\ref{g}) and (\ref{g0})] . For $\alpha _{L}=\alpha
_{R} $ and $|n-1|<1/3$, we obtain the maximum of the conductance remains
below the quantum of conductance times $\sin ^{2}(\pi n/2)$, neglecting the
effect of second derivatives of the real part of the self energy.

These terms disappear for $n=1$. This situation can be searched
experimentally adjusting the gate voltage in order to obtain the maximum
equilibrium conductance ($V\rightarrow 0$) . In this limit, our results
coincide with those obtained previously, using a different approach \cite{sela}. 
If in addition, the voltage drop is symmetric, the conductance up to
second order in $T$ and $V$ has a very simple expression \cite{note2}

\begin{eqnarray}
G &=&4\beta _{L}\beta _{R}\frac{2e^{2}}{h}[1-\frac{(\pi kT)^{2}}{3\widetilde{%
\Delta }^{2}}(1+2x^{2})  \nonumber \\
&&-\frac{(eV)^{2}}{4\widetilde{\Delta }^{2}}(1+5x^{2})].  \label{gu}
\end{eqnarray}

The ratio of the coefficient of $(eV)^{2}$ to that of $(kT)^{2}$ lies
between $3/(2\pi ^{2})=0.152$ in the strong coupling limit ($x=1$) to half
of this value in the non-interacting case ($x=0$). Reported values are 
$0.10\pm 0.015$ (Ref. \cite{grobis}), 0.15 \cite{kret} and $0.051\pm 0.01$ (Ref. \cite{scott}). 
The latter seems inconsistent with the predictions of the impurity
Anderson model.

{\it Note added}: After acceptance of this work we became aware of Ref. \cite{mun}, 
which addresses a similar problem for the particular case $\gamma=0$ ($\alpha_L=\beta_R$) 
using a perturbative approach in $\widetilde{\varepsilon }_{eff}^{\sigma }$. 
According to the results presented, this approach up to terms of total second order, 
leads to the same equations as in the Ng approximation \cite{note3,ng} 
[$\tilde{\Sigma}^{<}(\omega ) = 2i\tilde{f}(\omega ){\rm Im}\widetilde{\Sigma }^{r}(\omega )$], 
which trivially leads to the conservation of the current [see Eq. (\ref{dif})], 
but is unfortunately incorrect, as discussed in Section \ref{self}.

\ack

Helpful discussions with P. Roura-Bas, S. Florens, P. Schlottmann, and E.
Sela are gratefully acknowledged. The author is partially supported by
CONICET. This work was done in the framework of projects PIP No
11220080101821 of CONICET, and PICT R1776 of the ANPCyT, Argentina.


\begin{thebibliography}{99}
\bibitem{note3} The approximation 
$\Sigma^{<}(\omega )\simeq 2i\tilde{f}(\omega ){\rm Im}\Sigma^{r}(\omega )$
in this model is equivalent to
$G_{d \sigma}^{<}(\omega )\simeq 2i\tilde{f}(\omega ){\rm Im}G_{d \sigma}^{r}(\omega )$, 
as can be easily checked using Eqs. (\ref{gr}), (\ref{gl}) and the expression [Eq. (8) of Ref. \cite{none}]  
$g_{d\sigma }^{<}(\omega ) = 2i|g_{d\sigma }^{r}(\omega )|^{2}\sum_{\alpha
}\Delta _{\alpha }(\omega )f(\omega -\mu _{\alpha })$ for the nonintercating lesser Green's
function.
\bibitem{note1} From Eq. \ref{i}, neglecting the dependence of $\rho(\omega)$ with $V$, one obtains
$G \sim \rho(\alpha_L eV) + \rho(-\alpha_R eV)$. Thus, one expects peaks in $G$ for $eV$ near
$E_d/\alpha_L$, $(E_d+U)/\alpha_L$, $-E_d/\alpha_R$, and $-(E_d+U)/\alpha_R$.
\bibitem{note2} This result is also a simple extension of that given in Ref. \cite{ogu2}
for $\beta_L \neq \beta_R$.
\end{thebibliography}
\end{document}